# Dependence of nonthermal metallization kinetics on bond ionicity of compounds

R.A. Voronkov[1], N. Medvedev[2,3], A.E. Volkov[1,4,5,6]

[1]*P. N. Lebedev Physical Institute of the Russian Academy of Sciences, Leninskij pr., 53,119991 Moscow, Russia;*

[2]*Institute of Physics, Czech Academy of Sciences, Na Slovance 2, 182 21 Prague 8, Czech Republic;*

[3]*Institute of Plasma Physics, Czech Academy of Sciences, Za Slovankou 3, 182 00 Prague 8, Czech Republic;*

[4]*Joint Institute for Nuclear Research, Joliot-Curie 6, 141980 Dubna, Moscow Region, Russia;*

[5]*National Research Centre 'Kurchatov Institute', Kurchatov Sq. 1, 123182 Moscow, Russia;*

[6] *National University of Science and Technology MISiS, Leninskij pr., 4, 119049 Moscow, Russia*

**Abstract**

It is known that covalently bonded materials undergo nonthermal structure transformations upon ultrafast excitation of an electronic system, whereas metals exhibit phonon hardening. Here we study how ionic bonds react to electronic excitation. Density-functional molecular dynamics predicts that ionic crystals may melt nonthermally, however, into an electronically insulating state, in contrast to covalent materials. We demonstrate that the band gap behavior during nonthermal transitions depends on a bonding type: it is harder to collapse the band gap in more ionic compounds, which is illustrated by transformations in $Y_2O_3$ vs. NaCl, LiF and KBr.

**Keywords:** nonthermal phase transitions, nonthermal metallization, band gap dynamics, bond ionicity, ionic crystals

## 1. Introduction

Interaction of high-intensity femtosecond laser (e.g. free-electron laser, FEL) beams with matter results in a deposition of a high energy density into the electronic system of a target leading to nearly instantaneous increase of the electronic temperature up to several electron-Volts or higher [1,2]. Induced electron density redistribution causes changes in the interatomic potential that may trigger transitions between ordered material states or even disordering without significant lattice temperature increase – the so-called nonthermal melting [3,4].

A nonthermal structure transformation in an FEL spot occurs within sub-picosecond timescale (typically < 500 fs), i.e. at times shorter than those when thermal processes significantly heat up the lattice [5]. The latter ones are predicated on electron-lattice (electron-phonon) energy exchange and thus at near-threshold doses require a few picoseconds to take place [5].

Although nonthermal structure transformations have been known for a long time in the ultrafast laser community, only a few materials have been investigated yet (mostly elemental solids) because of experimental challenges and substantial computational efforts required for simulations [6–8]. General behavior patterns during nonthermal transitions have not been established yet.

It is known that metals show phonon hardening in the bulk upon femtosecond laser irradiation [9], whereas covalently bonded materials exhibit ultrafast nonthermal transformations. It has been suggested that ionic compounds may also become unstable upon electronic excitation [10]. However, pathways of these transformations differ from case to case resulting in necessity to simulate each material individually.

In contrast to elemental (one component) materials, compounds are not limited to covalent and metallic bond types. In this paper we consider two compound materials: $Y_2O_3$ and NaCl (and, to a lesser degree, LiF and KBr). Although both of these materials are usually considered as ionic crystals, the classification of bonding types is rather nominal and in many compounds one can speak only of a prevalence of a certain bonding type. Sodium chloride is a typical alkali-halide crystal with a strong ionic bonding, whereas yttria is a mixed-bonding transition metal oxide with a slight ionic type prevalence over the covalent one. The chosen materials allow us to study comparatively nonthermal effects in compounds with different levels of ionicity [11] and, taking into account previously investigated covalent materials, establish possible dependences of nonthermal transition pathways on the ionicity level of a material.

A comprehensive simulation of laser driven transformations requires multiscale models including quantum effects and hence often balances between an accuracy and a computational cost [5,12]. With this in mind, in this paper we do not consider processes of electronic system heating or relaxation. Since X-ray FELs deposit energy volumetrically into micrometric laser spots (both, in diameter and depth) [5], we neglect small temperature gradients and resulting slow energy sinks. We also neglect electron-phonon coupling that cools the electronic system and heats up the lattice, since it takes place on picosecond timescales, which are much longer than the times we analyze. This allows us to apply the density functional theory molecular dynamics (DFT-MD) [13] and focus purely on an effect of nonthermal transformations under an elevated electronic temperature.

In all investigated materials, we demonstrate a presence of nonthermal instability triggered by high electronic excitation. We estimate threshold electronic temperature or a deposited dose and a fraction of valence electrons excited to the conduction band that induces a nonthermal structure transformation within ~500 fs. We also analyze the band gap behavior during these transitions and find a temperature threshold of the band gap collapse in $Y_2O_3$, indicating a transition into a metallic state (here and further in the text by "insulating" and "metallic" we mean electronic conductivity), whereas no metallic state was possible to produce in NaCl, LiF and KBr.

## 2. Methods

We use density functional theory within the Quantum Espresso simulation package to study in detail nonthermal structure transformations in $Y_2O_3$ and NaCl [14]. After the geometry optimization, the initial lattice temperature was set to $T_i = 300$ K by equilibration of the kinetic and configuration temperatures via DFT molecular dynamics with the electronic system at zero temperature during 500 fs. Then, neglecting electron cascading which takes a few femtoseconds in a typical FEL spot except for hard X-rays [15], and does not significantly affect lattice dynamics [16], the electronic temperature was elevated. A series of molecular dynamics simulations within 500 fs with different electronic temperatures ranging from 1 eV to 6 eV were performed to identify structure transformation thresholds.

For certain electronic temperatures, simulation time was extended up to 1 ps. At such timescale electron-lattice interactions already may play a role, thus these simulations are used only to confirm a transition behavior in cases where this is unclear from 500 fs simulations.

We use norm-conserving pseudopotentials from the Quantum Espresso library and Perdew-Burke-Ernzerhof (PBE) exchange-correlation functional [17].

A simulated yttria supercell is composed of 2×2×2 cubic ($Ia\bar{3}$) primitive cells with 10 atoms in each, with the lattice parameter $a$=5.235 Å. A simulated sodium chloride supercell is composed of 2×2×2 cubic ($Fm\bar{3}m$) primitive cells with 8 atoms in each, with the lattice parameter $a$=5.615 Å.

The energy cutoff parameter controlling a size of the plane wave basis set was set as $E_{cut}$ ≈ 816 eV (60 Ry). During molecular dynamics simulations a single gamma point was used for calculations of forces, which is sufficient for simulation boxes of our sizes [18].

We use an NVT-ensemble (constant number of particles, volume and temperature) for the electronic system and NVE-ensemble (constant number of particles, volume and energy) for the atomic system since after an irradiation with FELs the unperturbed media is assumed to maintain a constant volume of the target's excited part in the bulk for times sufficiently longer than those modeled here. For all MD simulations the time step of 0.5 fs is used.

To identify a level of structure damage we calculated X-ray powder diffraction (XRD) patterns for each simulated electronic temperature with help of VESTA software [19].

## 3. Results

## 3.1 Y$_2$O$_3$

Figure 1 demonstrates XRD patterns of the supercell and corresponding atomic snapshots at the initial and final instants of simulations at the electronic temperatures of $T_e$ = 1.5 – 1.75 eV. At the electronic temperature of $T_e$ = 1.5 eV (corresponding to the deposited dose of 0.6 eV/atom), the atomic system is almost unperturbed. Little atomic displacements caused by interatomic potential modification in yttrium oxide occur at $T_e$ = 1.625 eV (the dose of 0.8 eV/atom, $n_e$ = 4.7% of electrons are excited to the conduction band). Significant damage within 500 fs appears only at $T_e$ = 1.75 eV (the dose of 1.0 eV/atom, $n_e$ = 5.4%). In this case, we can see in Figure 1 a noticeable decrease of the dominant diffraction peak and disappearance of the smaller peaks into the rising diffuse scattering background.

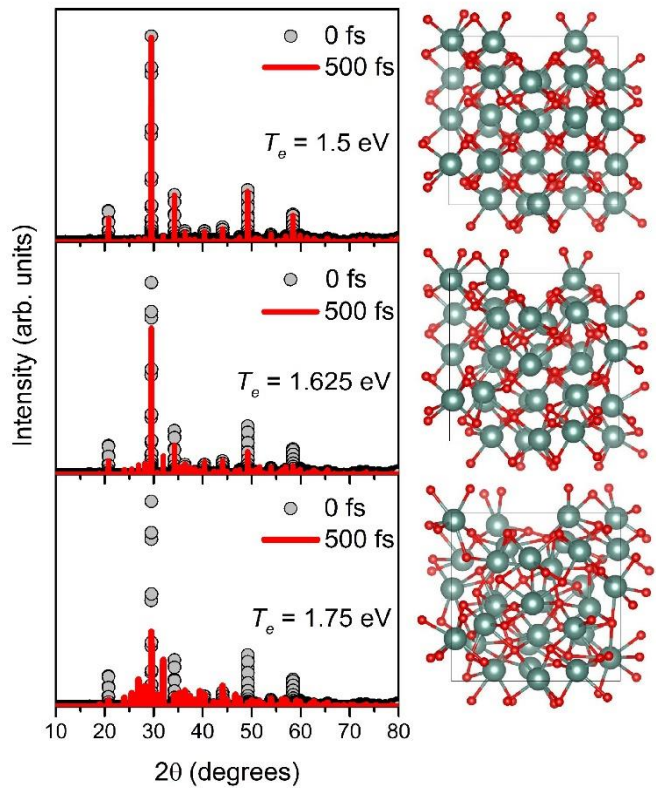

Figure 1. XRD patterns (λ=1.5406 Å) and corresponding atomic snapshots of the simulated yttria supercell at $T_e$ = 1.5 eV, $T_e$ = 1.625 eV and $T_e$ = 1.75 eV at the initial and final time instants.

To confirm that a structure modification at $T_e$ = 1.75 eV is indeed a nonthermal structure transformation rather than just strong atomic oscillations, we also calculated mean atomic displacements at each electronic temperature. Figure 2a shows that, indeed, continuous increase of atomic displacements at $T_e$ = 1.75 eV indicates disordering via diffusive behavior (displacements are proportional to the square root of time, see inset panel in Figure 2a), whereas a little structure perturbation at $T_e$ = 1.625 eV corresponds to strong atomic oscillations.

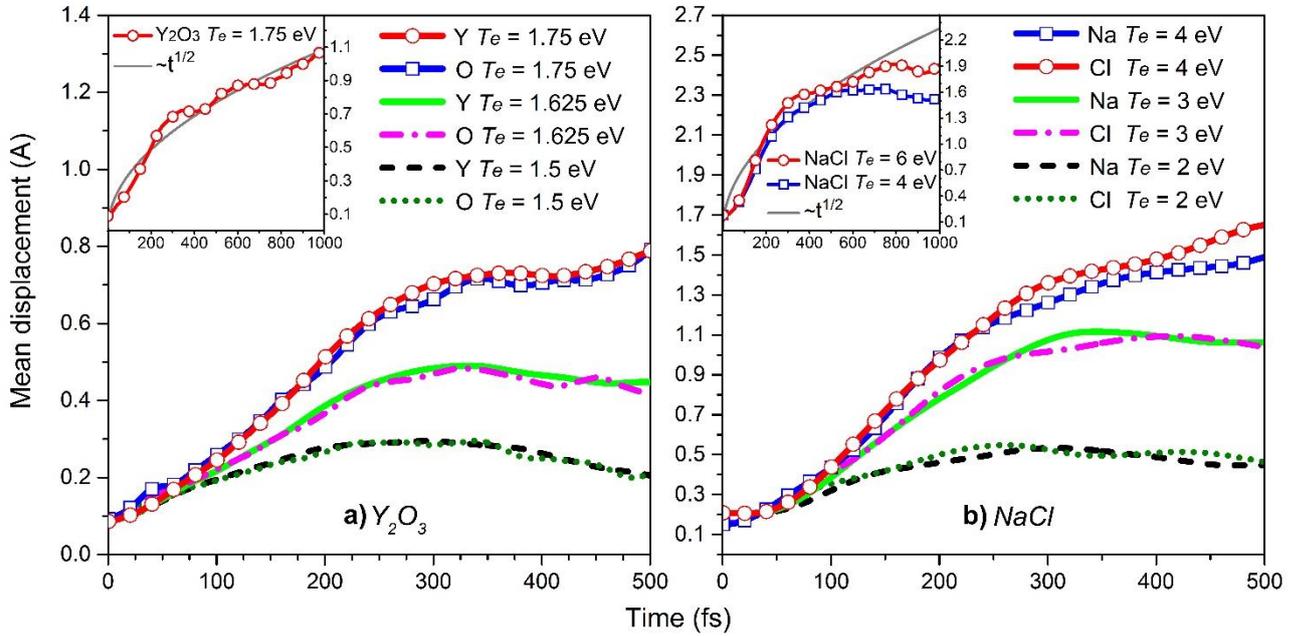

Figure 2. Mean atomic displacements in (a) yttria and (b) sodium chloride at different electronic temperatures. Insets represent displacements for a longer time interval.

In contrast to elemental covalent systems such as diamond and silicon [16,20], as well as group III-V semiconductors [21], nonthermal structure transformation at the threshold temperature in yttria is not accompanied by a band gap collapse. Instead, the band gap vanishes at a much higher electronic temperature of $T_e \geq 2.75$ eV (deposited dose of 3.6 eV/atom, $n_e$ = 11.3%) as shown in Figure 3a. Note that changes in the band gap at the initial time instant occur since the band structure depends on the electron occupation numbers (see e.g. [22] for details) whereas band gap oscillations around zero occur due to a finite number of atoms in the simulation box.

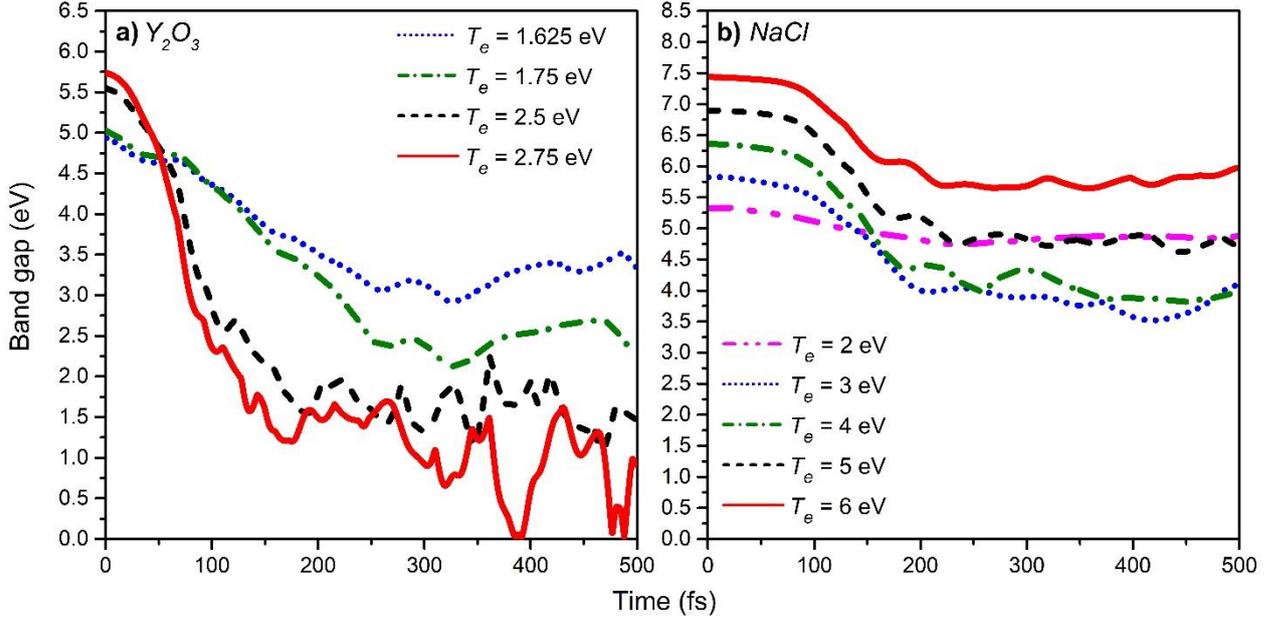

Figure 3. Evolution of the (a) yttria and (b) sodium chloride band gaps at different electronic temperatures.

Thus, at $T_e = 2.75$ eV a liquid-liquid state transition occurs in yttria turning it from a liquid insulating to a liquid metallic state. Such laser-driven state transitions have recently attracted physicists' attention as a promising mechanism for high-capacity memory devices construction [23].

Surprisingly, calculations within NPH (constant number of particles, pressure and enthalpy) ensemble for atoms demonstrate that yttria retains its initial structure up to $T_e \sim 2.5$ eV, although usually NPH nonthermal damage threshold is lower than the NVE one. This indicates that at doses below NPH damage threshold, at least at the sub-picosecond timescale after laser irradiation, finite-size yttria samples or near-surface regions may remain almost undamaged, whereas inside the bulk the nonthermal melting already occurs.

## 3.2 NaCl

In NaCl the first nonthermal damage occurs at $T_e = 2$ eV (deposited dose of 1.1 eV/atom, $n_e = 6.3\%$). According to XRD patterns in Figure 4, at higher doses structure modification becomes more significant. The main peak of the initial state at 32º becomes smaller than arising peaks of a damaged structure already at $T_e = 2.5$ eV (deposited dose of 2.1 eV/atom, $n_e = 9.6\%$) and vanishes completely at $T_e = 4$ eV (deposited dose of 7.2 eV/atom, $n_e = 19.5\%$). It seems that at $T_e = 6$ eV (deposited dose of 18.2 eV/atom, $n_e = 30.5\%$) the final state is amorphous. Peaks between 32º and 38º distinctively dominate over the scattering background but they form one broad peak characteristic of an amorphous state.

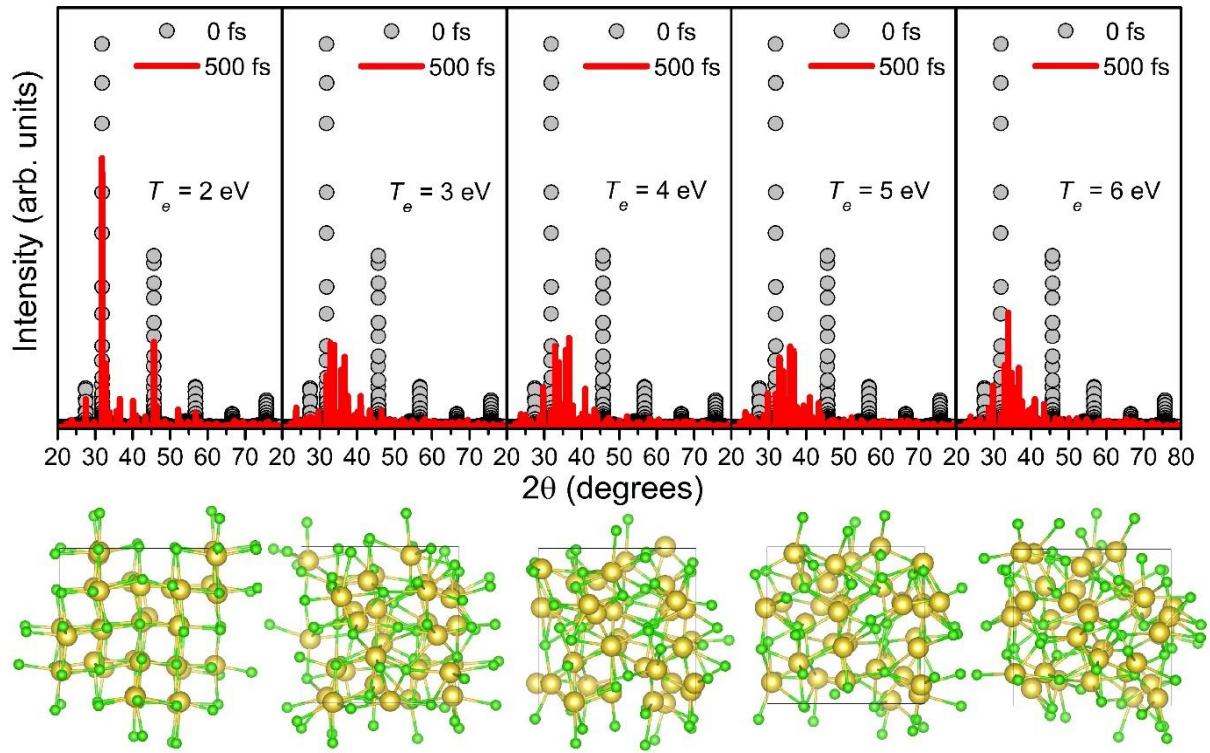

Figure 4. XRD patterns (λ=1.5406 Å) and corresponding atomic snapshots of the simulated sodium chloride supercell at different electronic temperatures at the initial and final time instants.

It is also clearly seen from Figure 2b – the mean atomic displacements saturate within 500 fs even at $T_e$ = 6 eV. The absence of a diffusive behavior in the atomic system means that at least up to this electronic temperature it is impossible to produce a liquid state in NaCl in the bulk via purely nonthermal melting.

In contrast to $Y_2O_3$ and other previously studied materials, the band gap in NaCl shrinks but does not collapse even up to $T_e$ = 6 eV (see Figure 3b). Thus, it appears not to be possible to produce a metallic state in NaCl by means of nonthermal melting up to such a high electronic temperature.

From the currently available data it follows that band gap behavior during nonthermal structure transformations depends on a bonding type. Indeed, in silicon, diamond and III-V compounds (with covalent bonding), the band gap collapse thresholds coincide with the corresponding nonthermal damage thresholds as reported in Refs. [20,21,24]. In $Y_2O_3$ and $Al_2O_3$ (see our previous work [25]), mixed-bonding crystals, the band gap collapses at doses high above their damage thresholds. Finally, in ionic NaCl, the band gap slightly shrinks but does not collapse at doses up to 7.2 eV/atom.

In order to check that NaCl is not an exception, we carried out similar calculations at $T_e$ = 6 eV for LiF and KBr. As one can see in Figure 5, at this electronic temperature, the band gaps of these materials shrink but do not collapse. Also, in contrast to NaCl and LiF, mean displacements in KBr demonstrate diffusive behavior indicating that there is no correlation between the band gap stability and a type of nonthermal structure transformation (solid-solid vs. solid-liquid). Thus, we conclude that, indeed, stability of the band gap in nonthermal transitions depends on ionicity level of a material.

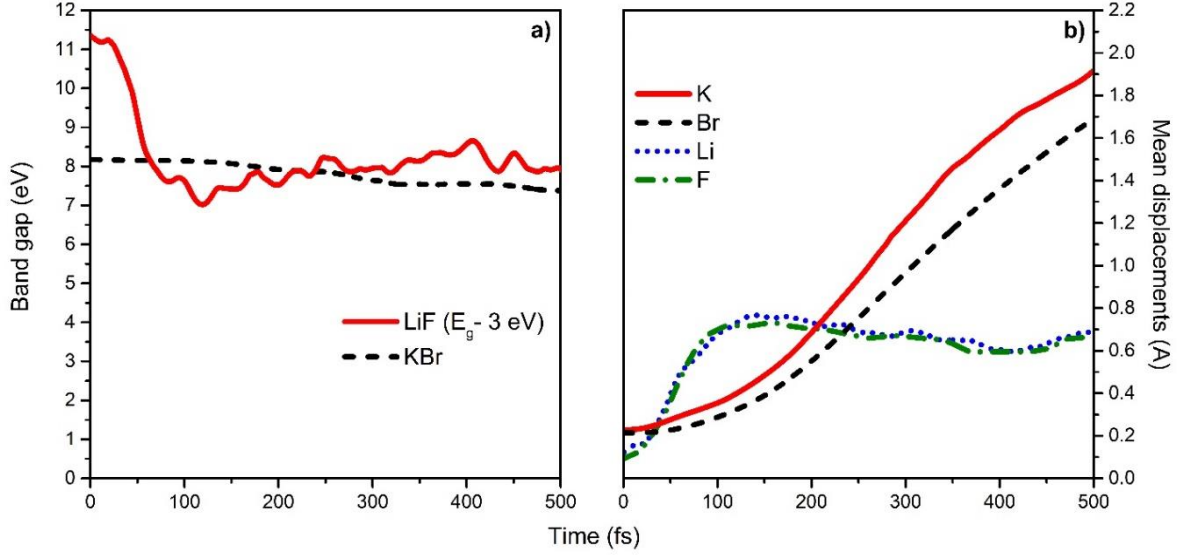

Figure 5. Evolution of band gaps (a) and mean atomic displacements (b)

in LiF and KBr at $T_e = 6$ eV.

## 4. Conclusions

We studied nonthermal structure transformations in a several ionic compounds: $Y_2O_3$, NaCl, LiF and KBr. It was demonstrated that an ultrafast nonthermal structure transformation occurs at electronic temperatures above $T_e = 1.75$ eV (dose of 1.0 eV/atom, 5.4% of electrons are excited to the conduction band) in $Y_2O_3$ and above $T_e = 2$ eV (deposited dose of 1.1 eV/atom, 6.3% of electrons excited) in NaCl.

In contrast to covalent-bonded materials, nonthermal structure transformations at threshold doses are not accompanied by the band gap collapse. Yttria turns into a metallic liquid at $T_e = 2.75$ eV (deposited dose of 3.6 eV/atom, 11.3% of excited electrons) within ~100-150 fs, whereas compounds with stronger ionic bondings – NaCl, LiF and KBr – remain electronically insulating during nonthermal transformations at least at electronic temperatures up to $T_e = 6$ eV.

We thus conclude that the band gap collapse caused by enhanced electronic temperatures depends on the level of ionicity in the material: more ionic crystals exhibit a larger and more robust band gap in electronically excited state, in comparison to more covalent ones. This finding open up a possibility for dynamically controllable band structure. By adjusting type and level of irradiation and selecting proper materials it may be potentially feasible to produce electronics tunable within fs-timescale. This conclusion should be validated in future dedicated experiments.

**Acknowledgements**

R.A.Voronkov acknowledges financial support of FAIR-Russia Research Center (FRRC). Partial financial support from the Czech Ministry of Education (Grants LTT17015 and LM2015083) is acknowledged by N. Medvedev. The work was supported by the Ministry of Science and High Education of the Russian Federation in the frameworks of Project No. 16 APPA (GSI) and Increase Competitiveness Program of NUST «MISiS»(Moscow, Russia) under Grant number K3-2018-041. The work of A.E. Volkov was supported by NRC Kurchatov Institute (Moscow, Russia) under Grant n.1603.

This work has been carried out using computing resources of the federal collective usage center Complex for Simulation and Data Processing for Mega-science Facilities at NRC "Kurchatov Institute", http://computing.nrcki.ru/. We also thank the GSI Helmholtzzentrum for providing the computational resources for this work.